\documentclass[preprint,12pt]{elsarticle}

\usepackage{graphicx}

\usepackage{amssymb}

\usepackage{blindtext, graphicx}
\usepackage{graphics}
\usepackage{amsmath}
\usepackage{color}
\usepackage{makeidx}
\usepackage{amssymb}
\usepackage{amstext}
\usepackage{url}
\usepackage{multirow}
\usepackage{color}
\usepackage[linesnumbered,ruled,vlined]{algorithm2e}
\usepackage{float}
\floatname{algorithm}{Procedure}
\newtheorem{mydef}{Definition}
\usepackage[table,xcdraw]{xcolor}
\usepackage{colortbl}
\usepackage[font=small,labelfont=bf]{caption}

\newenvironment{function1}[1][htb]
  {
   \begin{algorithm}
  }{\end{algorithm}}

\newenvironment{function2}[1][htb]
  {
   \begin{algorithm}
  }{\end{algorithm}}

  \usepackage{amsmath}

\newcommand{\ENCOD}{\texttt{EnCoD}}
\newcommand{\MENCOD}{\texttt{MEnCoD}}
\newcommand{\INV}{\mathbb{INV}}
\newcommand{\SIM}{\mathbb{SIM}}

\journal{Knowledge-Based Systems}

\begin{document}

\begin{frontmatter}


\title{Ensemble-based Overlapping Community Detection using Disjoint Community Structures}



\author{Tanmoy Chakraborty$^1$, Saptarshi Ghosh$^2$, Noseong Park$^3$\\
	$^1$ Dept. of CSE, IIIT Delhi, India\\
$^2$ Dept. of CSE, IIT Kharagpur, India\\
$^3$ Dept. of Information Sciences and Technology, George Mason University, USA\\
$^1$tanmoy@iiitd.ac.in, $^2$saptarshi@cse.iitkgp.ac.in, $^3$npark9@gmu.edu\\
{{\color{blue}(Accepted in Knowledge-Based Systems)}}
}

\begin{abstract}
While there has been a plethora of approaches for detecting
disjoint communities from real-world complex networks, some methods for detecting {\it overlapping} community structures have also been recently proposed. 
In this work, we argue that, instead of developing separate approaches for detecting overlapping communities, 
a promising alternative is to infer the overlapping communities
from multiple disjoint community structures.
We propose an ensemble-based approach, called \ENCOD, that leverages the solutions 
produced by various disjoint community detection algorithms to discover the 
overlapping community structure. 
Specifically, \ENCOD~generates a feature vector for each vertex from the results of  the base algorithms and learns which features lead to detect densely connected overlapping regions in an unsupervised way.
It keeps on iterating until the likelihood of each vertex belonging to its own community maximizes. 
{\color{black}Experiments on 
both synthetic and several real-world networks (with known ground-truth community structures) reveal that \ENCOD~significantly outperforms  nine state-of-the-art overlapping community detection algorithms.}  
Finally, we show that \ENCOD~is generic enough to be applied to 
networks where the vertices are associated with explicit semantic features.
To the best of our knowledge, \ENCOD~is the second {\em ensemble-based overlapping community detection approach} after MEDOC \cite{002PS16}.  
\end{abstract}


\end{frontmatter}


\section{Introduction}

{\color{black}Real-world networks are complex, high-dimensional and multi-faceted, thus can be interpreted in many different ways. 
A fundamental property of a real-world network is its ``community structure'', which is often assumed as organizational units in social networks~\cite{Zhang2015},
functional modules in biological networks~\cite{Palla},
scientific communities in citation networks~\cite{ChakrabortySTGM13}, and so on. 
Despite a huge amount of effort devoted in past decade or more~\cite{Fortunato201075,Xie:2013}, 
the problem of community detection (CD) has turned out to be more complicated because of two reasons -- 
(i)~CD is an {\em ill-defined} problem~\cite{Fortunato201075}; therefore one can obtain multiple solutions 
for a given network, each of which is important in its own way. 
Moreover a single objective function (even the ``best'', if exists) may not be able to effectively model such vast dimensions present in a network. 
(ii)~The vertices in real-world networks often belong to multiple communities~\cite{PalEtAl05,Leskovec}, leading to the community structure being overlapped. None of the existing disjoint CD algorithms are able to detect overlapping regions inside a community structure. 

Although a growing body of literature is focusing on discovering the overlapping community structure after the introductory work of Palla {\it et al.}~\cite{PalEtAl05},
the number is considerably less compared to the vast amount of literature on disjoint CD.
Moreover, in recent years, Chakraborty et al. \cite{1742,002PS16} showed that one can successfully detect overlapping regions inside a community structure by leveraging huge number of diverse and accurate  disjoint community structures.

In this paper, we attempt to utilize multiple views of the disjoint community structure observed in a given network (obtained from different disjoint CD algorithms) by  merging these views, 
distilling all their good qualities into an {\em ensemble solution}~\cite{DahlinS13,lanc12consensus}.
{\color{black} The motivation for our work comes from the fact that 
ensemble approaches have already been proved to be successful in clustering and classification tasks~\cite{Xu:2005}. Hence, ensemble approaches, if designed properly, can perform well in overlapping community detection as well.} 
However, the challenge in applying ensemble techniques to 
network-based analysis is due to the lack of sufficient information such as representative features of each vertex, extent of similarity between pair-wise vertices, 
and so on.

\begin{figure}[tb]
 \centering
 \includegraphics[width=\columnwidth]{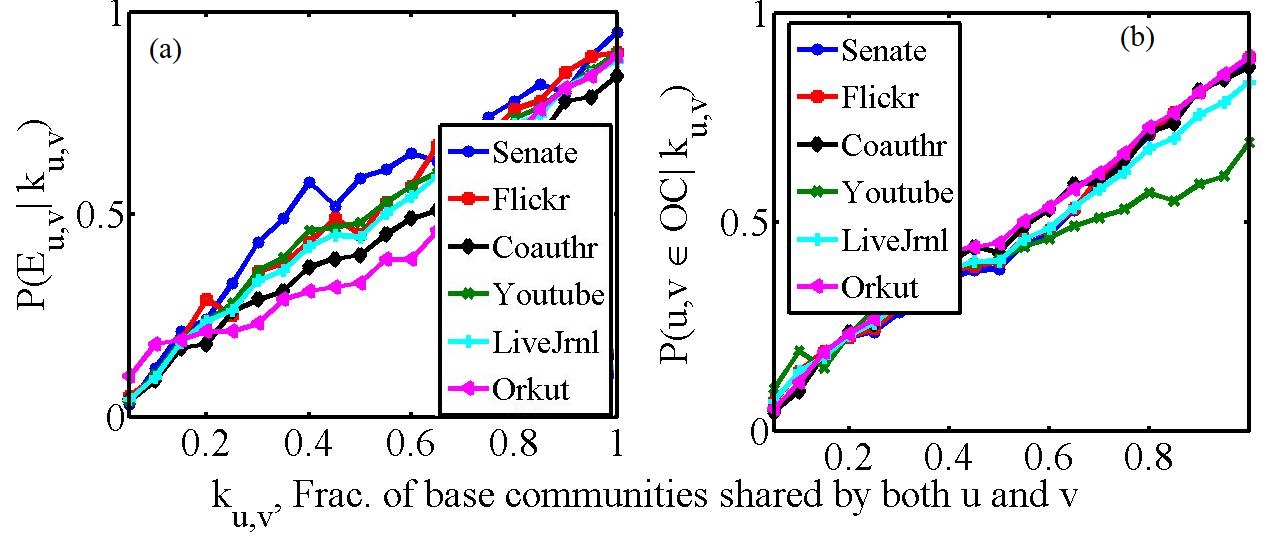}
 \caption{{\color{black}Conditional probability P (in y-axis)} of (a)~existence of an edge between two vertices, and (b)~common overlapping community (OC) membership of two vertices in the ground-truth community structure of six real networks, given the fraction of the base disjoint communities that both the vertices share. 
 We conclude that the more the vertices share common base disjoint communities, the more the chance that they are connected in the graph and belong to the same overlapping community.
\vspace{-5mm}
 }
 \label{evid}
\end{figure}

{\bf Empirical observations and Motivation.} Our algorithm is primarily  built on two interesting empirical observations on six large real networks (see Section~\ref{sec:setup} for dataset description). 
While analyzing the ground-truth community structures of these networks, {\color{black}we find many solutions by running each of the base disjoint CD algorithms on every single network. The solutions have significant (dis)similarity with the ground-truth, corroborating the observation in \cite{Chakraborty:2018}.}
We represent the memberships of each vertex in different base disjoint communities by a feature vector. 
Then we examine the probability of two vertices $u$ and $v$ being connected in the network, given that they have $k_{u,v}$ fraction of common base community memberships. 
Figure~\ref{evid}(a) plots this conditional probability for the six networks. We see an increasing relationship, i.e., {\em the more base disjoint communities a pair of vertices has in common, the higher the probability of an edge connecting them}. 

Further, we examine the conditional probability that two vertices belong to the same ground-truth overlapping community given that they have $k_{u,v}$ fraction of common base disjoint community memberships.
Once again we observe a positive correlation in Figure~\ref{evid}(b). 
Thus, {\em the probability of a pair of vertices belonging to the same overlapping community increases with the increase of their shared base community memberships}. 
The above observations motivated us to develop an overlapping CD
algorithm, by combining disjoint community structures identified for a given network.

{\bf Ensemble-based overlapping community detection.} 
Our method iteratively learns major dimensions of features for vertices that lead to densely-connected groups. 
In practice, since base disjoint CD algorithms produce many (and significantly different) community structures from a network, we leverage this information to extract the  latent feature information associated with each vertex.
\footnote{Note that the set of communities returned by a CD algorithm is termed as `community structure'.}
Our algorithm, named \ENCOD~is built on two hypotheses --
first, an overlapping community is formed by a group of densely-connected vertices; second, vertices within a community have significantly high feature similarity. 
We represent the memberships of each vertex in different base disjoint communities by a feature vector, and
then use the similarity between two vertices.
We then learn important features that may lead to discover densely-connected overlapping communities \cite{Li:2012,Li:2017}. 
To the best of our knowledge, \ENCOD~is the second algorithm after MEDOC \cite{002PS16} to propose {\em ensemble-based} overlapping CD algorithm by leveraging several disjoint community structures. Note that other two algorithms, namely  PEACOCK~\cite{Gregory2009} and PVOC~\cite{1742} can detect overlapping community from a {\em single} disjoint community structure by applying an efficient post-processing technique. They do not leverage {\em multiple} disjoint community structures.

 {\bf Summary of the evaluation.}
Experiments are conducted on both synthetic and six large real-world networks with known ground-truth overlapping community structure. We experiment with various disjoint CD algorithms as the base algorithms for \ENCOD. 
For comparative evaluation, we choose seven non-ensemble based overlapping CD algorithms. Additionally, we compare \ENCOD~with PVOC \cite{1742} and MEDOC \cite{002PS16}, two most recent overlapping CD algorithms that leverage disjoint community structures.
Few notable observations are as follows. \\
(i)~\ENCOD~is almost independent of how perfect the base CD algorithms are; however, stronger base algorithms have more impact in the aggregation stage than the weaker base algorithms. Also, the accuracy of \ENCOD~ never decreases with the increase in base solutions.\\
(ii)~\ENCOD~tends to get saturated after aggregating a certain number of base solutions, after which there is almost no effect on the final performance. However, the accuracy never decreases with the increase in base solutions. \\
(iii)~On average, \ENCOD~outperforms the best baseline overlapping CD algorithm in terms of the overlapping NMI metric ($ONMI$) \cite{journals}, $\Omega$ index and F-Score. \\
(iv)~We further apply \ENCOD~to networks where, apart from the topological structure, every vertex is associated with additional feature information. In such cases, \ENCOD~systematically combines the features into the model and produces more accurate overlapping community structure.    

{\color{black}
{\bf Contributions of the paper.} The major contributions of the paper are four-fold:
\begin{itemize}
    \item Ensemble algorithms for community detection have rarely been studied. This paper will pave the way for enhancing the state-of-the-art on ensemble based community detection.
    \item The idea of using the membership of a vertex in different disjoint community structures as features is unique and, to our knowledge, has not been used earlier in community detection.
    \item Empirical results indicate that \ENCOD\ is either as good as the best baseline CD algorithm, or sometimes performs even better than that. Therefore, one can rely on such ensemble framework instead of choosing and picking the best one from the jungle of CD algorithms available at present.
    \item The framework used in \ENCOD\ is generalized enough to be used to combine any feature set of vertices to detect community structure (as shown in Section~\ref{sec:vertexfeature}). Therefore, if the topology of a network is incomplete and sparse in nature and the properties of vertices are available, one can use the proposed framework to combine the features into the model for overlapping community detection. 
\end{itemize}
}
}

\section{Related Work}
Due to the abundance of literature on community detection (CD) algorithms~\cite{Fortunato201075,Xie:2013,Chakraborty:2017mca}, we restrict our discussion to some selected works that we deem as pertinent to our study. 
Two extensive survey articles on CD are~\cite{Fortunato201075} and \cite{Xie:2013}.

 {\bf Traditional community detection algorithms.} 
Most of the early research in CD assumed the communities to be disjoint. Many different CD approaches have been proposed, including, 
modularity optimization~\cite{blondel2008,Clauset2004,newman03fast},  
information theoretic approaches~\cite{Rosvall29012008}, 
vertex similarity-based approaches~\cite{JGAA-124}, significance-based approaches~\cite{oslom}, label propagation~\cite{labelprop},  diffusion-based approach~\cite{Raghavan:1057930},  
and so on (see~\cite{Fortunato201075,Xie:2013} for surveys).
Gradually, researchers realized that in real-world networks, 
a vertex can belong to multiple communities, resulting overlapping community structure. 
Palla {\it et al.} proposed `CFinder'~\cite{PalEtAl05}, the first ever method to detect overlapping communities based on clique-percolation technique. 
Subsequently, several overlapping CD techniques have been proposed, including modified modularity~\cite{chen2010detecting},
fitness function \cite{abs-1012-1269}, local density function \cite{BaumesGKMP05}, affinity propagation function \cite{ding}, integrated approach \cite{hajiabadi2017iedc}, node location analysis based approach \cite{zhi2016overlapping},
non-negative matrix factorization~\cite{Leskovec},  and so on. 
{\color{black}
Recently, Li et al. \cite{Li:2018:LSC} proposed a local spectral clustering for overlapping community detection. Huang et al. \cite{huang2018overlapping} proposed an overlapping community detection algorithm in heterogeneous social networks via the user model. Sattari and Zamanifar \cite{sattari2018spreading} suggested a spreading activation-based label propagation algorithm for overlapping community detection in dynamic social networks.}
Notably, though there have been several recent studies on the topic, the depth of study on overlapping community detection is 
much lesser than that for the disjoint case.

{\bf Ensemble-approach for community detection.} Ensemble approach has been well-studied in traditional data mining for clustering data points (see~\cite{Xu:2005} for a detailed review). 
These approaches can be classified into two categories \cite{Xu:2005}: object co-occurrence based approaches and median partitioning based approaches. 
However, this number is significantly less when we talk about clustering vertices in the network. A pioneering attempt was made by 
Dahlin and Svenson~\cite{DahlinS13} to propose ensemble clustering for network data.
Raghavan et al. \cite{PhysRevE.76.036106} showed the effectiveness of merging outputs of several community detection algorithms. Ovelg{\"{o}}nne and
Geyer{-}Schulz proposed CGGC \cite{OvelgonneG12}, an ensemble-based modularity maximization approach. 
Kanawati proposed YASCA \cite{Kanawati2014} that for each seed vertex considers its ego-centric network, partitions the network in different ways and combines the outputs. He further tuned two parameters -- quality and diversity of the partitions for final combination \cite{Kanawati2015}. 
Recently, Lancichinetti and Fortunato~\cite{lanc12consensus} proposed `consensus clustering' which utilizes a consensus matrix for disjoint community detection. 
However, all the above works focused on {\em identifying
disjoint communities}.

{\color{black}
Recently, two algorithms, PEACOCK~\cite{Gregory2009} and PVOC~\cite{1742} 
argued that intelligently post-processing a disjoint solution can lead to the detection of overlapping community structure. PVOC was reported to outperform PEACOCK \cite{1742}. 
 The difference between PVOC and \ENCOD\ is that PVOC uses a novel post-processing technique that modifies a given disjoint community structure using a metric called `permanence' \cite{Chakraborty:2014,AgarwalVAC18}, whereas \ENCOD\ does not do any post-processing; rather it considers multiple disjoint community structures and uses them as features of a vertex for detecting overlapping regions. Another ensemble algorithm, called MEDOC \cite{002PS16} uses the idea of meta community (community of communities). 
It creates a multi-partite network using the base disjoint community structure. After this, it runs an existing CD algorithm to partition the multi-partite network. Finally, a membership function is to determine the membership of a vertex to a community. \ENCOD\ does not create any meta-community using base community structures; rather it uses them to generate features for each vertex to be used further in the optimization algorithm.} In Section \ref{comparison}, we will show that our algorithm outperforms both PVOC and MEDOC.

\section{Ensemble-based Overlapping CD Algorithm}
\label{algo}

We design an overlapping CD algorithm by leveraging the solutions 
obtained from several disjoint CD algorithms.  This section details the proposed overlapping CD algorithm, named \ENCOD .
 We start by formally defining the ensemble-based overlapping community detection problem.

\noindent 
\begin{mydef}
{\bf (Ensemble-based Overlapping Community Detection)}   Given a network $G=(V,E)$ and a set of $M$ disjoint community detection algorithms $\mathcal{AL}=\{Al_m\}_{m=1}^M$, we aim at grouping the vertices into communities $\mathbb{OC}=\{OC_1, OC_2,\cdots\}$ in such a way that a vertex may belong to multiple communities and the following conditions are satisfied: $|\bigcup_{i} OC_i|=|V|$ and $\nexists i: OC_i = \phi$.
\end{mydef}

We now describe our algorithm, called \ENCOD,  an {\bf En}semble based Overlapping {\bf Co}mmunity {\bf D}etection. The overall framework of the algorithm is presented in Algorithm \ref{alg:medoc} and important notations are summarized in Table \ref{notation}. 

 {\bf Inputs to the algorithm:}
Two main inputs of the algorithm are the network $G(V,E)$ and the set of $M$ base algorithms ${\cal AL}=\{Al_m\}_{m=1}^M$ which yield disjoint community structures. Apart from these, one needs to specify: 
$K$, number of vertex orderings of a particular network on which a base CD algorithm will run;\footnote{Previous research reported that most disjoint CD algorithms produce different community structures depending on the ordering of vertices in the input~\cite{citeulike:12334819,Chakraborty:2014,lanc12consensus}.}
$\INV(v,C)$, an involvement function that will determine to what extent $v$ is a part of a base community $C$;
$\SIM'(C,v)$, a function that will measure the similarity of a vertex $v$ with a community $C$; and $\tau_L$, a threshold that confirms the minimum similarity value required by the vertices in a community to remain together, failing which the community is discarded. 
Possible definitions of these functions are discussed in Section~\ref{sec:parameter}. 

\begin{table}[!t]
 \centering
 \caption{Important notations used in this paper.}\label{notation}
  \vspace{-2mm}
 \scalebox{0.75}{
 \begin{tabular}{|c|p{1.1\columnwidth}|}
 \hline
 {\bf Notation} & \multicolumn{1}{c|}{{\bf Description}} \\\hline
 $G(V,E)$ & An undirected network with sets of vertices $V$ and edges $E$\\\hline
 $\mathcal{AL}$ & $\{Al_{m=1}^M\}$, set of $M$ base disjoint CD algorithms \\\hline
 $\mathbb{C}_m^k$ & $\{C_m^{1k},...,C_m^{ak}\}$, base community structure discovered by algorithm $Al_m$ on $k^{th}$ vertex ordering \\\hline
 $\Gamma_m$ & $\{\mathbb{C}_m^k\}_{k=1}^K$, set of  base disjoint community structures discovered  by base algorithm $Al_m$ on $K$ different vertex orderings\\\hline
 
 $\Gamma$ & $\Gamma_{m=1}^M$, set of all $MK$ base disjoint community structures\\\hline
 $\bar{a}$ & Average size of a base community structure\\\hline
 $\xi$& $\bar{a}MK$, approx. total number of communities in $\Gamma$ \\\hline
 $\mathbb{OC}$ & $\{OC_1,OC_2,\cdots\}$, final overlapping community structure consisting  of overlapping communities\\\hline
 $\mathbb{OC}_{t}$ & Overlapping community structure at iteration $t$ of \ENCOD\\\hline
  $\zeta$  & $\{\tau_1,\tau_2,\cdots\}$, set of thresholds where $\tau_j$ corresponds to overlapping community $OC_j$\\\hline
 $\tau_L$ & Global community threshold used to control the community size
 \\\hline
 $F_v$ & Feature vector of vertex $v$\\\hline
  
 \end{tabular}
 }
 \vspace{-5mm}

\end{table}

%

{\bf Summary of the algorithm:}
Motivated by the ego-centric circle detection algorithms proposed in \cite{Mcauley:2014,Chakraborty:2015}, we design \ENCOD~ that aims at maximizing the likelihood of the membership of each node inside a community. 
\ENCOD~ starts by running all the base disjoint CD algorithms on different vertex orderings\footnote{We here use random vertex ordering strategy \cite{citeulike:12334819}.} of a network, which produce dissimilar base community structures (Section \ref{sec:base}). 
Following this, the base community information is used to generate a feature vector for each vertex (Section \ref{sec:feature}); this in turn transforms the vertices in the network into a feature space. 
The algorithm now tries to infer the overlapping community structure from the vector representation of vertices, by iteratively optimizing an objective function (detailed in Section~\ref{sec:objective}). The actual iteration starts by assigning each vertex to a singleton community, and  a high similarity threshold value is assigned to each community to maintain high integrity among its constituent vertices. 
In each iteration, the algorithm manipulates the community structure by randomly removing vertices from their assigned communities and allocating vertices to some unassigned communities, such that the similarity condition is not violated (Section \ref{sec:manipulate}). 
After each iteration, the similarity threshold associated with each community is updated (Section \ref{sec:threshold}). The iteration continues as long as the value of the objective function does not decrease.

In the rest of the section, we elaborate the subroutines mentioned in Algorithm~\ref{alg:medoc}.

\begin{algorithm}[!h]
\scriptsize
\caption{{\ENCOD}: An {\bf En}semble based Overlapping {\bf Co}mmunity {\bf D}etection}
\label{alg:medoc}
\KwData{Network $G(V,E)$; \\A set of base algorithms ${\cal AL}=\{Al_m\}_{m=1}^M$; \\$K$: No of iterations; \\$\INV(.,.)$: Involvement function;\\ $\SIM'(.,.)$: Vertex-to-community similarity function;\\ $\tau_L$: Lower limit for community threshold}
\KwResult{Overlapping community structure $\mathbb{OC}$}

$\Gamma=\phi$ \\  

\For{each algorithm $Al_m\in {\cal AL}$}{
Generate $K$ different vertex orderings of $G$, run $Al_m$ on each vertex ordering and obtain $K$ disjoint community structures $\Gamma_m$. The size of each community structure $\mathbb{C}_m^k \in \Gamma_m$ is different. Each such set is denoted by $\mathbb{C}_m^k=\{C_m^{1k},...,C_m^{ak}\}$\label{algo1:ensemble}\\
$\Gamma=\Gamma \cup \Gamma_m$
}

\For{each $v\in V$}{
$F_v=ExtractFeatures(v,\Gamma, \INV)$;\\
}

\tcp{\color{blue}Initialization}
${\mathbb{OC}}=\phi$ \hfill \tcp{\color{blue}Final overlapping community structure}\label{oc}
${\mathbb{OC}}_0=\phi$ \hfill \tcp{\color{blue}Initial set of overlapping community }
${\zeta}_0=\phi$ \hfill \tcp{\color{blue}Initial set of community threshold}\label{zeta}
\For{each $v \in V$}{
$OC_0^v=\{v\}$\hfill \tcp{\color{blue}Each vertex is in each community}
${\mathbb{OC}}_0 = {\mathbb{OC}}_0 \cup OC_0^v$\\
$\tau_0^v=\infty$ \hfill \tcp{\color{blue}Arbitrary high threshold}
${\zeta}_0= {\zeta}_0 \cup \tau_0^v$
}

$Plogl=-1$ \hfill \tcp{\color{blue}Previous log likelihood}
$Clogl=-1$ \hfill \tcp{\color{blue}Current log likelihood}
$t=1$  \hfill \tcp{\color{blue}Current number of iterations}
${\mathbb{OC}}_t={\mathbb{OC}}_0$\\
${\zeta}_t= {\zeta}_0$\\

\While{$Clogl \geq Plogl$}{\label{cond}
$\mathbb{OC}={\mathbb{OC}}_t$\\
$Plogl=Clogl$\\
      ${\mathbb{OC}}_{t+1} = ManipulateComm(V,{\mathbb{OC}}_t)$\\     
     
     \tcp{\color{blue}Update thresholds and communities}
     \For{$OC_{t+1}^j \in {\mathbb{OC}}_{t+1}$}{
         $\tau_{t+1}^j \gets min \{\SIM'(OC_{t+1}^j,v) | v\in OC_{t+1}^j\}$ \label{thre}\\
	  \If{$\tau_{t+1}^j < \tau_L $ }{\label{thre_gl}\tcp{\color{blue}Threshold condition is violated}
		${\mathbb{OC}}_{t+1} \gets {\mathbb{OC}}_{t+1} \smallsetminus \{OC_{t+1}^j\}$\\
		  \For{each $v \in OC_{t+1}^j$}{
		      $\mathbb{OC}_{t+1} \gets \mathbb{OC}_{t+1} \cup \{v\}$
		   }
		${\zeta}_{t+1} \gets {\zeta}_{t+1} \smallsetminus \tau_{t+1}^j$\\
          }		
     }
  $Clogl= l_{ \zeta_{t+1}}(G;  \mathbb{OC}_{t+1})$\label{opt} [Eq. \ref{eq:loglikelihood}]   \\ 
  
$t=t+1$\\   
}
\Return $\mathbb{OC}$
\end{algorithm}

\subsection{Generate Base Communities} \label{sec:base}

 Assume that we are given a network $G=(V,E)$ and $M$ different base CD algorithms ${\cal AL}=\{Al_m\}_{m=1}^M$. For each such base algorithm $Al_m$, \ENCOD~ generates $K$ different vertex orderings of $G$ and runs $Al_m$ on each vertex ordering. This results in $K$ different community structures denoted as $\Gamma_m=\{\mathbb{C}_m^k\}_{k=1}^K$, where each community structure $\mathbb{C}_m^k=\{C_m^{1k},\cdots,C_m^{ak}\}$ constitutes a specific partitioning of vertices in $G$. The size of each $\mathbb{C}_m^k$ might be different. 
We choose to use different vertex orderings of $G$ since it has been shown
that many of the disjoint CD algorithms produce different community structures
depending on the ordering of vertices in the input~\cite{Chakraborty:2014,lanc12consensus}.
Here we use random vertex ordering strategy which ensures that the base algorithm starts from different seed vertices without any predefined bias and produces diverse community structures. At the end of this step, we obtain $\Gamma$, a set of $MK$ base community structures (Step \ref{algo1:ensemble} of Algorithm \ref{alg:medoc}).

 \begin{function1}[!t]\scriptsize
\caption{$ExtractFeatures(\text{Current vertex: } v,\text{Set of} \newline \text{base disjoint communities: } \Gamma, \text{Involvement function: } \INV)$}
 $F_v=\phi$; 
 $D_v=0$; 
 $\xi=0$; 
 $AF(v)=\phi$; \hfill \tcp{\color{blue}Auxiliary vector}
 \For{each $\Gamma_m \in \Gamma$}{\label{gammas} 
     \For{each $\mathbb{C}_m^k\in \Gamma_m$}{
      \For{each $C\in \mathbb{C}_m^k$}{\label{algo1:c}
       Measure $d_v^C=1-\INV(v,C)$;\label{algo1:dist} \\
       $AF_v=AF_v \cup \{d_v^C\}$;\label{algo1:fea}\\
       \If{$ d_v^C \geq D_v $}{
	    $D_v=d_v^C$;\label{algo1:D}
        }
        $\xi=\xi+1$;\label{algo1:cl}
       }      
   }
 }\label{gammae} 
 \For{each $AF_i(v) \in AF(v)$}{\label{algo1:pre}
     Compute $P(C_i|v) =\frac{D_v-AF_i(v) +1}{\xi.D_v + \xi -\sum_{k=1}^{\xi} AF_k(v)}$;\label{algo1:prob}\\
     $F_v=F_v\cup \{P(C_i|v)\}$;\label{algo1:post}
   }
   
\Return $F_v$   

\end{function1}

\subsection{Generate Feature Vectors for Vertices}\label{sec:feature}


The ensemble step in Function~1 (which is invoked in step 5 of Algorithm \ref{alg:medoc}) 
intends to leverage the base community structures $\Gamma$ to construct the feature vector of vertices.

\begin{mydef}
{\bf (Feature vector for a vertex)} Given a set of base community structures $\Gamma$,  the feature vector of vertex $v$ is denoted by $F_v$, whose $i^{th}$ entry $F_v(i)=P(C_i|v)$ indicates the probability of $v$ being a member of the base community $C_i$,  given vertex $v$ is observed. The length of $F_v$ is $|F_v|=\xi=\sum_{\Gamma_m\in \Gamma} \sum_{\mathbb{C}_m^k \in \Gamma_m}  |\mathbb{C}_m^k|$. The value of $P(C_i|v)$ is derived by measuring the involvement of $v$ into $C_i$.  
\end{mydef}

We now describe how we compute $F_v$ for a vertex v using Function~1.
To start with, we measure $\INV(v,C)$, the involvement $v$  into each base community $C\in\mathbb{C}_m^k$ (see Section \ref{sec:parameter} for the possible definitions of $\INV$). Note if $v$ is not a part of a certain community $C$, we leave the corresponding field as zero. The  value returned by $\INV$ is subtracted from $1$ to get the  distance of $v$ from the community  (Step \ref{algo1:dist}). Following this, we construct an auxiliary feature vector $AF(v)$ for vertex $v$, which consists of elements indicating the distance of $v$ from each base community (Step \ref{algo1:fea}). The number of communities $\xi\in \Gamma$ can be approximated by $\xi=\bar{a}MK$, where the average size of a community is denoted by $\bar{a}$. Therefore, the size of $AF(v)$ is same as $\xi$. $D_v=\max_i AF_i(v)$ denotes the maximum distance of $v$ from any of the base community present in $\Gamma$ (Step \ref{algo1:D}). 

To construct the {\em actual feature vector} $F_v$ for vertex $v$, we measure the probability associated with each base community $C_i$ given that vertex $v$ is observed. For a given vertex $v$, the community label $C_i$ is assumed to be a random variable from a distribution with probabilities $\{P(C_i|v)\}_{i=1}^{\xi}$. One can provide a non-parametric estimation of such probabilities based on the evidences of the data and the community structure. 
To transfer the distance into probability (the smaller the distance, the larger the probability), we use the following conditional probability \cite{002PS16} for vertex $v$ to be a part of $C_i$  (Step~\ref{algo1:prob}):
\begin{equation}
 P(C_i|v) =\frac{D_v-AF_i(v) +1}{\xi.D_v + \xi -\sum_{k=1}^{\xi} AF_k(v)}
\end{equation}
The denominator of the above equation 
ensures that $\sum_{i=1}^{\xi} P(C_i|v)=1$. 
Adding 1 in the numerator ensures a non-zero probability for all vertex-community pairs. 
For smaller distance values, $P(C_i|v)$ increases proportionately to $D_v-AF_i(v)$, i.e., the larger the deviation of $AF_i(v)$ from $D_v$, the more the increase. As a consequence, the corresponding community $C_i$ becomes more likely for $v$. 
Now we can construct the actual feature vector $F_v$ for  $v$ as $F_v=\{ P(C_k|v) \}_{k=1}^ {\xi}$ (Step \ref{algo1:post}).
This process essentially maps a vertex to a $\xi$-dimensional vector space.

\subsection{Objective Function}\label{sec:objective}


After the ensemble step, each vertex is represented by a feature vector. Our method of detecting the overlapping community structure (denoted by $\mathbb{OC}=\{OC_1,OC_2,\cdots\}$) aims at maximizing a certain objective function which represents the likelihood of each vertex being a part of its own community. The likelihood function assumes that if two vertices $u$ and $v$ are found to have very similar feature vectors, then they -- (i) should be connected by an edge in the network, and (ii) should stay in the same community. These two assumptions are supported by the observations in Figure \ref{evid}.  
We use the knowledge of the edges in the network to validate the inferred community structure.

To start with, let us define few measures required to formulate the objective function.  Note that all the communities mentioned in this subsection correspond to the overlapping communities that we aim to detect from a network. 


\begin{mydef} \label{dis}
{\bf Similarity between two vertices.}
For vertices $u$ and $v$, let $F_u=\{F_u^i\}_{i=1}^{\xi}$ and $F_v=\{F_v^i\}_{i=1}^{\xi}$ represent their feature vectors respectively. Then the similarity between two vertices $u$ and $v$ is calculated using the cosine similarity among the feature vectors.
 \begin{equation}
 \label{eq:sim1}
 \SIM(u,v) =  \frac{F_u.F_v}{|F_u||F_v|}= \frac{\sum_{i=1}^{\xi} F_u^iF_v^i}{ \sqrt{ \sum_{i=1}^{\xi} {F_u^i}^2} \sqrt{ \sum_{i=1}^{\xi} {F_v^i}^2 }} 
\end{equation}
\end{mydef}
We also tried other similarity measures (see Section~\ref{sec:parameter}), and obtained best results with cosine similarity. With this similarity measure, we further define the similarity of a vertex with a community as follows.

\begin{mydef}\label{disc}
 {\bf Similarity between a vertex and a community.} Given a community $OC$ and a vertex $v$, the similarity of $v$ and $OC$ is calculated as the average similarity of $v$ with all other vertices in $OC$.
 \begin{equation}
  \label{eq:sim2}
  \SIM'(OC,v)=\frac{\sum_{u\in OC} \SIM(u,v)}{|OC|}
 \end{equation}
\end{mydef}
 
Each community is associated with a minimum similarity threshold which every vertex needs to satisfy in order to be a part of the community.

\begin{mydef}\label{def:thre}
{\bf Similarity threshold per community.} Each community $OC_j$ in our model is assigned a similarity threshold $\tau_j$ in such a way that  if vertex $v \in V$ is in $OC_j$ then it should satisfy $\SIM'(OC_j,v) \geq \tau_j$.  
\end{mydef}

Given $\mathbb{OC}=\{OC_1,OC_2,\cdots\}$, an overlapping community structure and a set of thresholds $\zeta=\{\tau_1,\tau_2,\cdots\}$, we then define a {\em membership similarity} between two vertices based on their membership in different overlapping communities.

\begin{mydef}
{\bf Membership similarity of pair-wise vertices.} 
The membership similarity of two vertices $u$ and $v$ is defined in terms of their memberships in different communities as follows:
\begin{equation}
\beta(u,v) = \exp\{{[\beta_1(u,v)}]^2 - {[\beta_2(u,v)]}^2\}
\end{equation}
\end{mydef}
where, 
\begin{equation}
 \begin{split}
 \beta_1(u,v) = \sum_{OC_j: \{u,v\} \subseteq OC_j} {(\SIM(u,v) - \tau_j + \lambda)}^{-1}\\
\beta_2(u,v) = \sum_{OC_j: \{u,v\} \nsubseteq OC_j} {(\SIM(u,v) - \tau_j + \lambda)}^{-1}
 \end{split}
\end{equation}

 We consider $\{u,v\} \subseteq OC_j$ if both the vertices $u$ and $v$ belong to the same community $OC_j$; whereas $\{u,v\} \nsubseteq OC_j$  if $OC_j$ does not contain either $u$ or $v$ or both. 
One should consider $\lambda$ as large as possible so that individual terms in the summation are always positive. In practice, we set $\lambda$ as the maximum of all the threshold values corresponding to the communities mentioned in Definition \ref{def:thre} (i.e., $\lambda=max\{\tau_1,\tau_2,\cdots\}$). The value of $\beta_1(u,v)$ ({\em resp.} $\beta_2(u,v)$) tends to become high if both $u$ and $v$ share ({\em resp.} do not share) common communities with very high thresholds. 
Furthermore, the value of membership similarity $\beta(u,v)$ depends not only on 
the common community memberships, but also on the threshold specific to the communities. 
In particular, with the increase of the number of the common communities where both $u$ and $v$ belong to and the community threshold, the value of $\beta(u,v)$ also increases.
 
Given the membership similarity of all pair-wise vertices, we  further define the probability that a pair of vertices $u$ and $v$ are connected by an edge $E_{u,v}$ as follows:
\begin{equation}
\label{eq:prob}
p(E_{u,v}\in E) = \frac{\beta(u,v)}{1+\beta(u,v)}
\end{equation}
Similarly, the probability of two vertices {\it not} being connected is $p(E_{u,v}\notin E)=1-p(E_{u,v}\in E)$. The value of $p(E_{u,v}\in E)$ increases with the increase in $\beta(u,v)$, and add-one smoothing is applied to $p(E_{u,v}\in E)$ in order to avoid ``divided by zero'' situation. One can thus obtain the predicted probability $p(E_{u,v}\in E)$ from the information of the overlapping community structure $\mathbb{OC}$ and the set of thresholds $\zeta$. 
The idea is that, the more the similarity between the feature vectors of two vertices, the common communities both of them are part of and the similarity threshold of the common communities, the more the membership similarity between them.  Given the probability estimation, our model should therefore assure that its resultant network corresponds to the real network.

 Let us assume that each edge is generated independently. Then the joint probability of $G$ and $\mathbb{OC}$ 
 is estimated as:
\begin{equation}
\label{eq:probgraph}
P_{\zeta}(G; \mathbb{OC}) = \prod_{E_{u,v} \in E} p(E_{u,v} \in E) \prod_{E_{u,v} \notin E}p(E_{u,v} \notin E)
\end{equation}
The log likelihood of the joint probability is:
\begin{equation}
\small
\begin{split}
l_{\zeta}(G; \mathbb{OC}) &=  \log{(P_{\zeta}(G; \mathbb{OC}))} \\
& = \sum_{E_{u,v} \in E} \log{(p(E_{u,v} \in E))} + \sum_{E_{u,v} \notin E} \log{(p(E_{u,v} \notin E))}\\
 & = \sum_{E_{u,v}\in E}\log{(\beta(u,v))} - \sum_{E_{u,v} \in V\times V}\log(1+\beta(u,v))\\ 
&  = \sum_{(u,v)\in E}\phi(u,v) - \sum_{E_{u,v} \in V\times V}  \log(1+ \exp\{\phi(u,v)\})  
\end{split}
\label{eq:loglikelihood}
\end{equation} 
where $\phi(u,v) =  \log{(\beta(u,v))} = ({[\beta_1(u,v)}]^2 - {[\beta_2(u,v)]}^2) $

\ENCOD~attempts to maximize the objective function mentioned in Equation \ref{eq:loglikelihood} (see line \ref{opt} in Algorithm \ref{alg:medoc}) in order to obtain the final overlapping community structure $\mathbb{OC}$. 

\begin{function2}\scriptsize
\caption{$ManipulateComm(\text{Set of vertices: } V,\text{Current}\newline \text{overlapping community structure at iteration $t$: } {\mathbb{OC}}_t)$}
${\mathbb{OC}}_{t+1} = {\mathbb{OC}}_{t}$\\
\For{each $v$ in $V$}{
${S1}_t^v = \{OC_t^j | OC_t^j \in {\mathbb{OC}}_{t} \wedge v \in OC_t^j\}$\\
${S2}_t^v = \{OC_t^j | OC_t^j \in {\mathbb{OC}}_{t} \wedge v \notin OC_t^j\}$\\
\tcp{\color{blue}$Rand(x,y)$: a random number between $x$ and $y$}
$p_1 \gets Rand(1,|{S2}_t^v|) $\\
$AC_{t+1}^v \gets  \left \lceil{ \frac{p_1 + |{S1}_t^v|}{|{S1}_t^v|} } \right \rceil$\\

$p_2 \gets Rand(1,|{S1}_t^v|)$\\
$RC_{t+1}^v \gets \left \lceil{ \frac{p_2 + |{S1}_t^v|}{|{S1}_t^v|} } \right \rceil$\\
\text{Choose randomly} $OC_{AC} \text{ such that } OC_{AC} \subseteq {S2}_t^v,|OC_{AC}| = AC_{t+1}^v$\\
\For{each $OC_t^j \in OC_{AC}$}{
$OC_{t+1}^j \gets OC_{t+1}^j \cup \{v\}$\\
}
\text{Choose randomly} $OC_{RC} \text{ such that }   OC_{RC} \subseteq {S1}_t^v,|OC_{RC}| = RC_{t+1}^v$\\
\For{each $OC_t^j \in OC_{RC}$}{
$OC_{t+1}^j \gets OC_{t+1}^j\smallsetminus \{v\}$
}
}
\Return ${\mathbb{OC}}_{t+1}$

\end{function2}


\subsection{Manipulate Community Structure}\label{sec:manipulate}

The framework for manipulating the overlapping community structure in each iteration is shown in Function 2. The function originally starts with the initializations mentioned in lines \ref{oc}-\ref{zeta} of Algorithm 1. Initially at iteration $t=0$, each vertex $v$ is assigned to a separate community $OC_0^v$ with a very high threshold value $\tau_0^v$. In every iteration, we change the memberships of $v$ by removing it randomly from certain communities it belongs to, 
and assigning it to certain communities where it did not belong to. 
The threshold corresponding to each community is updated accordingly such that the constraint in Definition~\ref{def:thre} is not violated. For a certain community $C_j$, if threshold $\tau_j$ falls below a certain lower limit $\tau_L$, the community is discarded because such a community is too heterogeneous to represent a dense group in our model. 
The value of $\tau_L$ is empirically determined as described in Section~\ref{sec:parameter}. 
Since we intend to discover the communities of each vertex, 
in every iteration, we concentrate more on those vertices which are yet assigned to less number of communities, than those vertices which are already assigned to many communities.
Therefore, with the increase in the number of communities a vertex is already part of after iteration $t$, it is less likely that the community membership of $v$ is manipulated at iteration $t+1$.

At each iteration $t$, the function $ManipulateComm$ takes the set of overlapping communities $\mathbb{OC}_t=\{OC_t^1, OC_t^2,\cdots\}$. Then for each vertex $v$, ${S1}_t^v = \{OC_t^j | OC_t^j \in {\mathbb{OC}}_{t} \wedge v \in OC_t^j\}$ indicates the set of communities where $v$ belongs to at iteration $t$. Similarly, ${S2}_t^v = \{OC_t^j | OC_t^j \in {\mathbb{OC}}_{t} \wedge v \notin OC_t^j\}$ indicates those set of communities where $v$ is currently not a member of. 

Now, let $AC_{t+1}^v$ denote the number of communities to add $v$ to, and $RC_{t+1}^v$ denote the number of communities to remove $v$ from. These variables are measured as follows: 
\begin{eqnarray}
AC_{t+1}^v \gets  \left \lceil{ \frac{p_1 + |{S1}_t^v|}{|{S1}_t^v|}} \right \rceil , \; 
RC_{t+1}^v \gets \left \lceil{ \frac{p_2 + |{S1}_t^v|}{|{S1}_t^v|} }\right \rceil
\end{eqnarray}
Here, {\color{black}$p_1$ is a random integer chosen from $[1,|{S2}_t^v|)$ such that the number of communities $AC_{t+1}^v$ to add $v$ to, never exceeds $|{S2}_t^v|$, indicating the number of $v$'s current communities.
Similarly, $p_2$ is an integer randomly chosen from $[1, |{S1}_t^v|)$ such that the value of $RC_{t+1}^v$ is less than or equal to $|{S1}_t^v|$ (number of communities that $v$ is currently part of).} 
Note both the values $AC_{t+1}^v$ and $RC_{t+1}^v$ tend to decrease with the increase of $|{S1}_t^v|$, ensuring that, the more the number of communities $v$ is currently part of, the less the manipulation of $v$ in the current iteration, and vice versa.  

After selecting the number of communities $v$ is to be added to and to be removed from,  we choose the set of communities $OC_{AC}$ from  ${S2}_t^v$ randomly such that $|OC_{AC}|=AC_{t+1}^v$ and assign $v$ to $OC_{AC}$ (Step 12). Similarly, we randomly choose $RC_{t+1}^v$ many communities (indicated by the set $OC_{RC}$) from ${S1}_t^v$ and remove $v$ from it (Step 14).  The corresponding changed communities are updated accordingly.

\subsection{Update Thresholds and Compute Objective Function}\label{sec:threshold}

Once the manipulation of the current overlapping community structure is over (as described above), we obtain a new overlapping community structure $\mathbb{OC}_{t+1}$. 
The next task is to update the threshold corresponding to the new overlapping community structure (Steps 23-31 in Algorithm 1).  For community $OC_{t+1}^j \in \mathbb{OC}_{t+1}$, we set the new threshold $\tau_{t+1}^j$ as the minimum value of all the similarities of the $OC_{t+1}^j$'s constituent vertices  with $OC_{t+1}^j$ itself, i.e., $\tau_{t+1}^j = min \{\SIM'(OC_{t+1}^j,v) | v\in OC_{t+1}^j\}$ (Step \ref{thre}). This assignment ensures that Definition \ref{def:thre} is never violated. 

Apart from the minimum threshold assigned to each community, we also maintain another global threshold $\tau_L$, which ensures the overall coherence among the members in each community. Once all the community-centric thresholds are updated, we further check whether all these 
thresholds are higher than $\tau_L$ (Step~\ref{thre_gl}). If some threshold  $\tau_{t+1}^j$ corresponding to community $OC_{t+1}^j$ falls below $\tau_L$, we discard $OC_{t+1}^j$ and set each of its constituent members as a separate community (Steps 26--29). 
The value of $\tau_L$ controls the size of the final overlapping communities -- a higher value of $\tau_L$ results in smaller community size. The selection of the value of $\tau_L$ is described later
in Section~\ref{sec:parameter}.

Finally, we compute the log likelihood after the current iteration  
$l_{\zeta_{t+1}}(G; \mathbb{OC}_{t+1})$ using Equation~\ref{eq:loglikelihood} (Step~\ref{opt}) and compare it with $l_{\zeta_{t}}(G; \mathbb{OC}_{t})$ (Step~\ref{cond}). 
If the former is greater than the latter, we keep the current community structure $\mathbb{OC}_{t+1}$ and  the current set of thresholds $\zeta_{t+1}$
and continue the iteration. 
Otherwise, we discard the updated sets and retain the sets obtained in the previous iteration (Step 21). We terminate the process when there is no sufficient increase of the log likelihood.
Note that this step is not shown in Algorithm 1 (which rather shows that the iteration 
immediately stops once the condition is violated) as it depends on the users' confidence on the number of iterations she wants to continue after the log likelihood condition is violated. 
In practice, continuing the iterations ensures that the algorithm does {\it not} get stuck
at any local optima of the objective function.

\subsection{Complexity Analysis}

\ENCOD~terminates when it reaches a maxima of the objective function and the objective function does not change further. In the worst case, after any iteration the maximum number of overlapping communities is $|V|$ and the maximum number of vertices within each community is also $|V|$. Therefore, the runtime after each iteration is $\mathcal{O}(|V|+|\mathbb{OC}|)$. 
Importantly, a manipulation of the current community structure is only possible if the log likelihood increases. Therefore, \ENCOD~generally converges after a finite number of iterations. In practice, we assume that the local maxima is reached if the log likelihood does not change after    $|V|$ iterations.

Note that, since \ENCOD~ is an ensemble algorithm, it requires the running of all base algorithms. 
However, speedup might be achieved by parallelizing the base algorithms, or by using fewer base algorithms (potentially at the cost of performance, as analyzed later in Section~\ref{sec:result}).

\begin{table*}[!ht]
	\caption{{\color{black}Description of the real-world networks ($\tau$: avg. edge-density per community, $\bar{c}$: avg. community size, $V_c$: avg. number of community memberships per vertex)}.}
	\label{dataset}
	
	\centering
	\scalebox{0.8}{
		\begin{tabular}{l||l|l|l|c}
			\hline
			Network & Vertex type & Edge type & Community type  & Reference\\\hline
			Senate & Senate & Similar voting pattern & Congress & \cite{PhysRevE.91.012821}\\ 
			Flickr & User & Friendship & Joined group &  \cite{Wang-etal12} \\
			Coauthorship & Researcher & Collaborations & Publication venues &   \cite{Palla}\\
			Youtube & User & Friendship & User-defined group & \cite{Leskovec} \\
			LiveJournal & User & Friendship & User-defined group  &  \cite{Leskovec} \\
			Orkut  & User & Friendship & User-defined group  & \cite{Leskovec}\\\hline
	\end{tabular}	}
\vspace*{-3mm}
\end{table*}

\scalebox{0.8}{
	\begin{tabular}{l||l|l|l|r|r|r}
			\hline
			Network &  \# vertices & \# edges  & \# communities & \multicolumn{1}{l|}{$\tau$} & \multicolumn{1}{l|}{$\bar{c}$} & \multicolumn{1}{l}{$V_c$} \\\hline
			 Senate & 1,884 & 16,662 & 110    & 0.65 & 81.59 & 4.76 \\ 
			 Flickr & 80,513 & 5,899,882 & 171 & 0.046 & 470.83 & 18.96  \\
			Coauthorship & 391,526 & 873,775 & 8,493 & 0.231 & 393.18& 10.45 \\
			 Youtube & 1,134,890 &	2,987,624 & 8,385 & 0.732 & 43.88 & 2.27 \\
			 LiveJournal & 3,997,962 & 34,681,189 & 310,092 & 0.536  & 40.02 & 3.09 \\
			Orkut & 3,072,441 & 117,185,083 & 6,288,363 & 0.245 & 34.86 & 95.93\\\hline
	\end{tabular}}

\section{Experimental Setup}  \label{sec:setup}

\noindent
In this section, we describe the datasets and the experimental setup used to evaluate the
performance of \ENCOD{}.

\subsection {Datasets}  
We use two types of network datasets to evaluate various CD algorithms:
\noindent (1)~{\bf Synthetic networks:} We use the LFR benchmark model~\cite{PhysRevE} to generate synthetic networks along with their ground-truth communities.
We follow the parameters configuration suggested in~\cite{1742} while generating networks:
number of vertices $N=10,000$, average degree $\bar k=50$, maximum degree $k_{max}=150$, maximum community size $c_{max}=150$, minimum community size $c_{min}=50$,  percentage of overlapping vertices $O_n=15\%$, 
number of communities to which a vertex belongs $O_m=20$, and
mixing parameter $\mu=0.3$ (representing the ratio of inter- and intra-community edges; the lower the value of $\mu$, the better the communities are).
For each configuration, we create $100$ such LFR networks and report the average result.
\noindent (2)~{\bf Real-world networks:} We use six real-world networks of widely different scales, whose ground-truth communities are available {\em a priori} (see Table~\ref{dataset}).

\subsection { Evaluation Metrics}   
To compare the detected overlapping community structure with the ground-truth, we use three standard metrics: 
(1)~Overlapping Normalized Mutual Information ($ONMI$)~\cite{journals}, 
(2)~Omega ($\Omega$) Index~\cite{Leskovec}, and (3)~F-Score~\cite{Leskovec}.  

\if{0}
The availability of ground-truth communities allows us to quantitatively evaluate the performance of community detection algorithms. For
evaluation, we use metrics that quantify the level of correspondence between the detected and the ground-truth communities. Given a network
$G(V, E)$, we consider a set of ground-truth communities $C$ and a set of detected communities $C^*$ where each ground-truth community $C_i
\in C$ and each detected community $C_i^* \in C^*$ is defined by a set of its member vertices. To quantify the level of correspondence of $C$
to $C^*$, we consider the following three validation metrics.

\noindent $\bullet$ {\bf Omega ($\Omega$) Index.} It is the accuracy measures on estimating the number of communities that each pair of nodes shares \cite{Gregory}:
\begin{equation}
 \frac{1}{|V|^2} \sum_{u,v\in V} 1 \{|C_{uv}| = |C^*_{uv}|\}
\end{equation}

where $C_{uv}$ is the set of ground-truth communities that $u$ and $v$ share, and $C^*_{uv}$ is the set of detected communities that 
$u$ and $v$ share.

\noindent $\bullet$ {\bf F-Score.}  It measures the correspondence between each detected community to one of the ground-truth communities \cite{Leskovec}. To compute it, we
need to determine which $C^*_i \in C^*$ corresponds to which $C_i \in C$. We define F-Score to be the average of the F-Score of the
best-matching ground-truth community to each detected community, and the F-score of the best-matching detected community to
each ground-truth community as follows:
\begin{equation}
 \frac{1}{2}(\frac{1}{|C|} \sum_{C_i \in C} F(C_i,C^*_{g(i)})   +   \frac{1}{|C^*|} \sum_{C^*_i \in C^*} F(C_{g^{'}(i)},C^*_i)   )
\end{equation}
where the best matching $g$ and $g^{'}$ is defined as follows: $g(i)=\operatorname*{arg\,max}_j F(C_i, C^{*}_{j})$, 
$g^{'}(i)=\operatorname*{arg\,max}_j F(C_j, C^{*}_{i})$, and $F(C_i,C^*_j)$ is the harmonic mean of Precision and
Recall.

\noindent$\bullet$ {\bf Overlapping Normalized Mutual Information \\(ONMI).} It is an information theory based  criterion to compare the
detected communities and the ground-truth communities (see \cite{journals} for details). We use the ONMI implementation written by the authors\footnote{\small It is available at
\url{https://github.com/aaronmcdaid/Overlapping-NMI}}.
\fi

\subsection { Base Disjoint CD Algorithms for Aggregation:}   
There exist numerous disjoint CD algorithms, which differ in the way they define the community structure. Here
we select the following  disjoint CD algorithms from different categories based on their working principles: 
(i) {\em Modularity-based approach}: FastGreedy \cite{newman03fast}, Louvain  \cite{blondel2008} and CNM \cite{Clauset2004}; 
(ii) {\em Node similarity-based approach}: WalkTrap \cite{JGAA-124}; 
(iii) {\em Compression-based approach}: InfoMap  \cite{Rosvall29012008}; 
(iv) {\em Diffusion-based approach}: Label Propagation \cite{labelprop}. 
\ENCOD~ ensembles the outputs of all these six algorithms 
and generates the feature vectors for the vertices.

\subsection {\bf Baseline Algorithms:}  
We compare \ENCOD~ with the following seven non-ensemble based overlapping CD algorithms that cover different types of overlapping CD heuristics:  
 OSLOM~\cite{oslom}, EAGLE~\cite{Shen},  COPRA~\cite{Gregory1}, SLPA~\cite{Xie}, 
 MOSES \cite{moses}, BIGCLAM~\cite{Leskovec} and {\color{black}IEDC \cite{hajiabadi2017iedc}}.
We further compare \ENCOD~ with PVOC~\cite{1742} and MEDOC \cite{002PS16} which are the most recent algorithms that use disjoint community structure to detect the overlapping communities.

\section{Experimental Results} \label{sec:result}

\noindent In this section, we discuss how the performance of \ENCOD~depends on various parameters, 
and compare its performance with state-of-the-art overlapping CD algorithms.
We run \ENCOD~on both synthetic and real-world networks, and compare its performance with that of seven state-of-the-art overlapping CD algorithms.

\begin{figure}[!t]
 \centering
 \includegraphics[width=0.7\columnwidth]{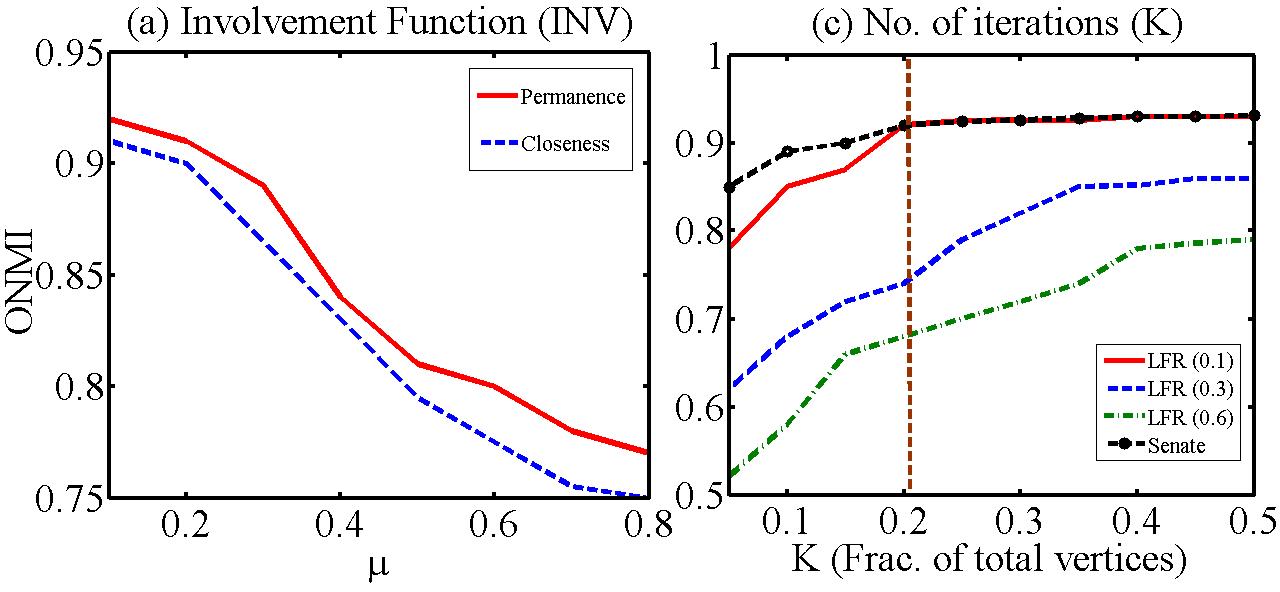}
 \includegraphics[width=0.7\columnwidth]{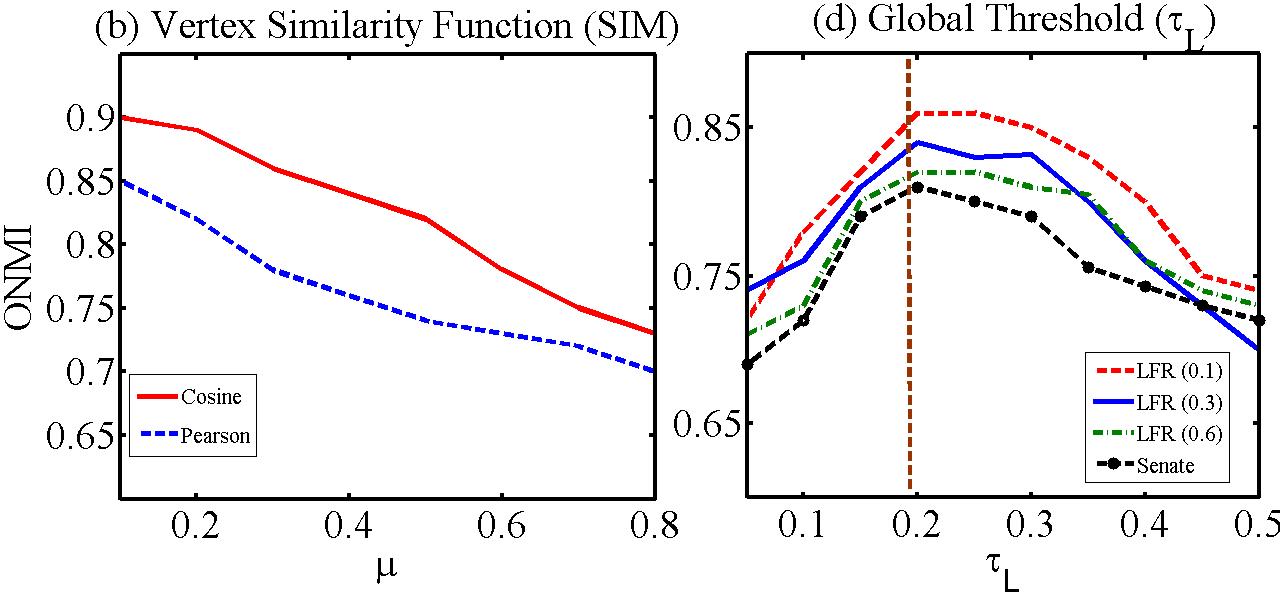}
 \caption{Change in accuracy of \ENCOD~ with different values of the algorithmic parameters. We vary $\mu$ (the mixing parameter of LFR) from $0.1$ to $0.8$. 
 In (c) and (d), we vary $K$ and $\tau_L$ respectively and report the accuracy for three different LFR and Senate networks. 
}
\label{parameter}
\vspace*{-5mm}
\end{figure}

\subsection{Parameter Selection} \label{sec:parameter}

\noindent For parameter selection in \ENCOD, we perform experiments 
over LFR networks of widely-varying quality of ground-truth community structures, 
obtained by varying the mixing parameter $\mu$ from $0.1$ to $0.8$ with the increase of $0.1$.

 \noindent $\bullet$ {\bf Involvement Function ($\mathbb{INV}$):} The following two functions are used to quantify what extent a vertex is involved in a community \cite{002PS16}: 
(i)~{\em Closeness Centrality} measures how close vertex $v$ is from other vertices where $v$ is a part of, i.e., $\frac{|C|}{\sum_{u \in C} dist(u,v)}$, where $dist(u,v)$ is the shortest distance from $u$ to $v$; 
(ii)~{\em Permanence}~\cite{Chakraborty:2014} measures how permanent is vertex is in its own community, i.e., $Perm(v,C) = \frac{I(v,G[C])}{D(v) E_{max}(v)} - (1-c_{in}(v,G[C]))$, where $G[C]$ is the induced subgraph of $C$, $D(v)$ is the total degree of $v$, $I(v,G[C])$ is the internal degree of $v$ w.r.t. $G[C]$, $E_{max}(v)$ is the maximum external connections to any one of the external communities of $v$, and $c_{in}(v,G[C])$ is the clustering coefficient of $v$ w.r.t. $G[C]$ (see more in \cite{Chakraborty:2014})\footnote{\small We also checked the performance of the algorithm by simply assigning $1$ to the entry of $v$'s feature vector if it belongs to the corresponding community, and $0$ otherwise (without measuring its involvement). However, we did not obtain good results.}.
Figure~\ref{parameter}(a) shows that the performance of \ENCOD~ tends to be always superior with $\INV$ as permanence. 

 \noindent $\bullet$ {\bf Similarity between two vertices ($\mathbb{SIM}$):} Although in Definition \ref{dis}, we have mentioned to measure the similarity between the feature vectors of two vertices using cosine similarity, we also measure the similarity with  {\em Pearson correlation coefficient} between two feature vectors. Cosine similarity turns out to be the better similarity measure than Pearson correlation coefficient 
 (see Figure~\ref{parameter}(b)).

\noindent $\bullet$ {\bf Number of iterations ($K$):} We set $K$ as a function of the number of vertices $N$ in the network. The analysis in Figure \ref{parameter}(c) confirms that for most of the networks, especially the ones which possess distinct community structure  (such as LFR network with $\mu = 0.1$, Senate), 
the performance of \ENCOD~ converges at $K = 0.2|V|$. We therefore consider this value as default.

 \noindent $\bullet$ {\bf Global threshold ($\tau_L$):} $\tau_L$ controls the size of the overlapping communities -- the size changes inversely with the value of $\tau_L$. We vary $\tau_L$ from $0.05$ to $0.50$ (in steps of $0.05$) and observe that the maximum accuracy is attained at $\tau_L=0.20$ for most of the networks (Figure~\ref{parameter}(d)). 
 
\noindent Therefore, unless otherwise stated, we will report the results of \ENCOD~ with the following parameter settings: 
$\INV=permanence$, $\SIM=cosine$, $K = 0.2|V|$ and $\tau_L=0.20$.

\begin{figure}[tb]
 \centering
 \includegraphics[width=\columnwidth]{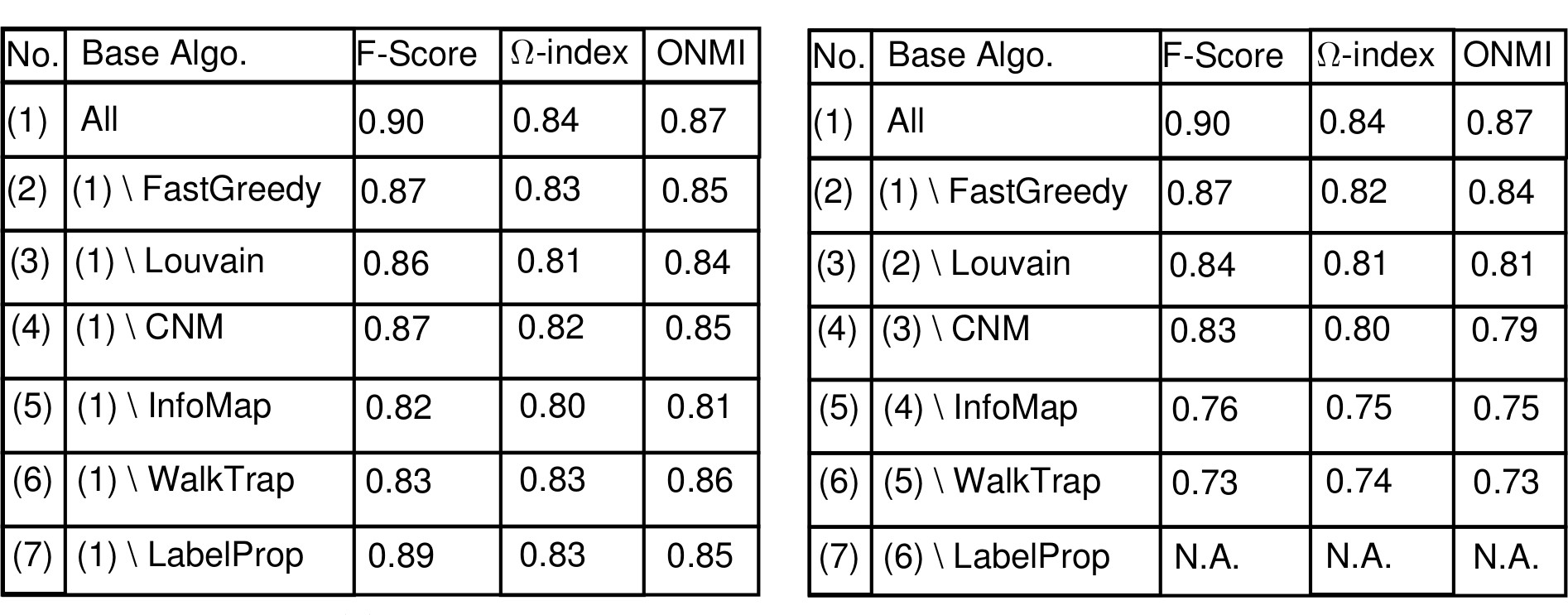}
 \vspace*{-5mm}
 \caption{Effect of base CD algorithms: performance of \ENCOD~after (a)~ removing base algorithms in isolation, i.e., one at a time, (b)~removing base algorithms one after another. When all base algorithms are removed, \ENCOD~does not work (N.A.).}
\label{runtime}
 
\end{figure}

\subsection{Comparative Evaluation}\label{comparison}

 Here we perform a two-fold evaluation of $\ENCOD$ with the baseline algorithms.

{\color{black}
\noindent \textbf{Comparison with state-of-the-art overlapping CD algorithms.}
We run \ENCOD~and seven other state-of-the-art overlapping CD algorithms on default LFR network and six real-world networks, and compare the output with the ground-truth community structures in terms of three evaluation metrics. 
Tables \ref{tab:compare-onmi}, \ref{tab:compare-omega}, and \ref{tab:compare-fscore} show the performance of the competing methods in terms of ONMI, $\Omega$ Index and F-Score respectively. 

Overall, \ENCOD~ shows the best performance, followed by BIGCLAM.
The absolute average ONMI of \ENCOD{} for all  the networks taken together is $0.82$, which is $14.08$\% higher than BIGCLAM ($0.71$). The absolute average values of  $\Omega$ index and F-Score for \ENCOD{} are $0.80$ and $0.80$ respectively. }


\begin{table}[t]
\centering
\caption{{\color{black}Performance (ONMI) of the competing algorithms w.r.t. the ground-truth community structure of LFR and six real-world networks.}}
\label{tab:compare-onmi}
\scalebox{0.8}{
\begin{tabular}{|l|c|c|c|c|c|c|c|}
\hline
\multirow{2}{*}{\textbf{Algorithm}} & \multicolumn{7}{c|}{\textbf{ONMI}}                                                                                                   \\ \cline{2-8} 
                                    & \textbf{LFR}  & \textbf{Senate} & \textbf{Flickr} & \textbf{Coauth} & \textbf{Youtube} & \textbf{LiveJ} & \textbf{Orkut} \\ \hline
\textbf{OSLOM}                      & 0.217       & 0.798           & 0.158           & 0.546                 & 0.410             & 0.429                & 0.436         \\ \hline
\textbf{COPRA}                      & 0.870          & 0.688          & 0.197          & 0.624                 & 0.393            & 0.413                & 0.504          \\ \hline
\textbf{SLPA}                       & 0.87          & 0.579          & 0.276          & 0.585                 & 0.639            & 0.607                & 0.798          \\ \hline
\textbf{MOSES}                      & 0.783         & 0.512          & 0.395           & 0.391                  & 0.369            & 0.388                & 0.537         \\ \hline
\textbf{EAGLE}                      & 0.174         & 0.714           & 0.158           & 0.195                 & 0.492            & 0.712                & 0.714          \\ \hline
\textbf{BIGCLAM}                    & 0.791         & 0.728           & 0.678           & 0.654                 & 0.705            & 0.701                & 0.741          \\ \hline
\textbf{IEDC} & 0.771 & 0.732 & 0.453 & 0.398 & 0.384 & 0.356 & 0.587 \\\hline
\textbf{\ENCOD}                      & \textbf{0.87} & \textbf{0.84}   & \textbf{0.79}   & \textbf{0.78}         & \textbf{0.82}    & \textbf{0.81}        & \textbf{0.84}  \\ \hline
\end{tabular}}
\end{table}

\begin{table}[t]
\centering
\caption{{\color{black}Performance ($\Omega$ Index) of the competing algorithms w.r.t. the ground-truth community structure of LFR and six real-world networks.}}
\label{tab:compare-omega}
\scalebox{0.8}{
\begin{tabular}{|l|c|c|c|c|c|c|c|}
\hline
\multicolumn{1}{|c|}{\multirow{2}{*}{\textbf{Algorithm}}} & \multicolumn{7}{c|}{\textbf{$\Omega$ Index}}                                                                             \\ \cline{2-8} 
\multicolumn{1}{|c|}{}                                    & \textbf{LFR}  & \textbf{Senate} & \textbf{Flickr} & \textbf{Coauth} & \textbf{Youtube} & \textbf{LiveJ} & \textbf{Orkut} \\ \hline
\textbf{OSLOM}                                            & 0.735         & 0.656           & 0.10             & 0.592           & 0.415            & 0.395          & 0.563          \\ \hline
\textbf{COPRA}                                            & 0.472         & 0.656           & 0.15            & 0.592           & 0.674            & 0.641          & 0.717          \\ \hline
\textbf{SLPA}                                             & 0.577         & \textbf{0.82}   & 0.15            & 0.592           & \textbf{0.83}    & \textbf{0.79}  & \textbf{0.82}  \\ \hline
\textbf{MOSES}                                            & 0.525         & 0.765           & 0.25            & 0.691           & 0.622            & 0.592          & 0.563          \\ \hline
\textbf{EAGLE}                                            & 0.787         & 0.765           & 0.20             & 0.543           & 0.570            & 0.543          & 0.666          \\ \hline
\textbf{BIGCLAM}                                          & 0.735         & 0.765           & 0.667           & 0.657           & 0.708            & 0.706          & 0.701          \\ \hline
\textbf{IEDC} & 0.765 & 0.743 & 0.565 & 0.435 & 0.421 & 0.452 & 0.613 \\\hline
\textbf{\ENCOD}                                            & \textbf{0.84} & \textbf{0.82}   & \textbf{0.75}   & \textbf{0.79}   & \textbf{0.83}    & {\bf 0.79}           & \textbf{0.82}  \\ \hline
\end{tabular}}
\end{table}

\begin{table}[t]
\centering
\caption{{\color{black}Performance (F-Score) of the competing algorithms w.r.t. the ground-truth community structure of LFR and six real-world networks.}}
\label{tab:compare-fscore}
\scalebox{0.8}{
\begin{tabular}{|l|l|l|l|l|l|l|l|}
\hline
\multicolumn{1}{|c|}{\multirow{2}{*}{\textbf{Algorithm}}} & \multicolumn{7}{c|}{\textbf{F-Score}}                                                                                   \\ \cline{2-8} 
\multicolumn{1}{|c|}{}                                    & \textbf{LFR} & \textbf{Senate} & \textbf{Flickr} & \textbf{Coauth} & \textbf{Youtube} & \textbf{LiveJ} & \textbf{Orkut} \\ \hline
\textbf{OSLOM}                                            & 0.112        & 0.579           & 0.387           & 0.48            & 0.25             & 0.259          & 0.421          \\ \hline
\textbf{COPRA}                                            & 0.506        & 0.526           & 0.72            & 0.48            & 0.55             & 0.570          & 0.421         \\ \hline
\textbf{SLPA}                                             & 0.562        & 0.316           & \textbf{0.886}  & 0.64            & 0.35             & 0.311          & 0.632          \\ \hline
\textbf{MOSES}                                            & 0.562        & 0.474           & 0.498           & 0.48            & 0.35             & 0.415          & 0.474          \\ \hline
\textbf{EAGLE}                                            & 0.112        & 0.474           & 0.498           & 0.426           & 0.45             & 0.466          & 0.579          \\ \hline
\textbf{BIGCLAM}                                          & 0.731        & 0.692           & 0.626           & 0.707           & 0.720            & 0.754          & 0.718          \\ \hline
\textbf{IEDC} &0.821 & 0.703 & 0.598& 0.667& 0.489& 0.549& 0.705 \\\hline
\textbf{\ENCOD}                            & \textbf{0.90} & \textbf{0.79}   & 0.72            & \textbf{0.80}    & \textbf{0.80}     & \textbf{0.83}  & \textbf{0.79}  \\ \hline
\end{tabular}}
\end{table}

\subsection{Effect of Base Algorithms} \label{sub:base-algo-effect}

\noindent One can anticipate some noise in the ensemble process due to those base CD algorithms which are generally weak in detecting accurate communities. Therefore we test which base algorithm has the most effect on the final output of \ENCOD{}. 
To this end, we first consider all base algorithms together and measure the performance of \ENCOD. 
Subsequently, we remove each base algorithm in isolation (i.e., one at a time) and compare the change in accuracy due to the removal.  
Figure~\ref{runtime}(a) shows that for all the removals, the accuracy decreases significantly.
InfoMap (which has been shown to be the best among all the base algorithms~\cite{Rosvall29012008}) seems to be the one for which the accuracy of \ENCOD{} deteriorates significantly. 
But Label Propagation and WalkTrap seem to have less effect on the performance of \ENCOD. 

In another experiment, we remove the base algorithms one after another (not in isolation), and
report the performance of \ENCOD{} in Figure \ref{runtime}(b).
Even with only the two `weakest' base CD algorithms (WalkTrap and Label Propagation), \ENCOD{} achieves more than 80\% of its performance when all six base algorithms are used.


In general, these results indicate that combining results from all sorts of (strong or weak) base CD 
algorithms  is important to enhance the final performance of the ensemble method.
Importantly, using more base algorithms never decreases the performance of the ensemble method.

{\color{black}
 {\bf Comparison with PVOC and MEDOC.} As mentioned earlier, 
\ENCOD~is the fourth algorithm after PEACOCK~\cite{Gregory2009},  PVOC~\cite{1742} and MEDOC \cite{002PS16} which use disjoint community structure to detect the overlapping communities\footnote{Recall that MEDOC is an ensemble algorithm that uses multiple disjoint community structures, whereas PEACOCK and PVOC apply a post-processing on a single disjoint community structure to detect overlapping regions}. 
PVOC has been proved to outperform PEACOCK \cite{1742}; 
hence we compare \ENCOD~with PVOC along with MEDOC and present the results in Table~\ref{pvoc}. 
For all the networks, \ENCOD~performs significantly better than both PVOC and MEDOC.  The performance of \ENCOD~(PVOC, MEDOC) averaged over all the datasets is as follows -- ONMI: $0.82$ ($0.74,0.75$), $\Omega$ Index: $0.81$ ($0.75,0.70$), F-Score: $0.80$	($0.68,0.76$).}

{\color{black}
The superior performance of \ENCOD\ over PVOC and MEDOC is potentially due to the following reason. In a sense, the final performance of both PVOC and MEDOC depends upon the performance of a single CD algorithm. PVOC  brings in the overlapping property as a post-processing step after finding a single disjoint community structure, and hence the quality of the overlapping community structure is dependent upon that of the disjoint community structure found initially. 
MEDOC uses a CD algorithm to partition the multi-partite network that is created using the base disjoint community structures; hence the quality of the final overlapping community structure depends on the performance of the CD algorithm used over the multi-partite network.
On the other hand, \ENCOD~obtains the features of vertices from the disjoint community structures given by several disjoint CD algorithms, and uses these features in an optimization setting. Thus the performance of \ENCOD~does not depend upon that of any one CD algorithm, and it can effectively learn from multiple disjoint community structures.

}

\begin{table}[tb]
 \centering
 \caption{Comparison of \ENCOD~with PVOC and MEDOC on different networks.}\label{pvoc}
  \scalebox{0.6}{
 \begin{tabular}{|l|c|c|c||c|c|c||c|c|c|}
 \hline
 
\multirow{2}{*}{{\bf Network}} & \multicolumn{3}{c||}{{\bf PVOC}} & \multicolumn{3}{c||}{{\bf MEDOC}} &  \multicolumn{3}{c|}{\ENCOD} \\\cline{2-10}
        & {\bf F-Score}  & {\bf $\Omega$ Index} & {\bf ONMI} &  {\bf F-Score}  & {\bf $\Omega$ Index} & {\bf ONMI} & {\bf F-Score}  & {\bf $\Omega$ Index} & {\bf ONMI}\\\hline
{\bf LFR}	& 0.75 & 0.78 & 0.79 & 0.80 & 0.78 & 	0.82 & 0.90 & 0.84 & 0.87\\
{\bf Senate} &  0.69 & 0.73 & 0.71 & 0.78 & 0.80 & 0.76 & 0.79 & 0.82 & 0.84\\
{\bf Flickr} & 0.62 & 0.65 & 0.67 & 0.70 & 0.73 & 0.71 & 0.72 & 0.75 & 0.79\\
{\bf Coauthorship} & 0.64 & 0.68 & 0.69 & 0.74 & 0.71 & 0.70 & 0.80 & 0.79 & 0.78\\
{\bf Youtube} &  0.63 & 0.63 & 0.67 & 0.70 & 0.66 & 0.64 & 0.80 & 0.83 & 0.82\\
{\bf LiveJournal} & 0.74 & 0.72 & 0.71 & 0.80 & 0.76 & 0.78 & 0.83 & 0.79 & 0.81\\
{\bf Orkut} & 0.69 & 0.72 & 0.75 & 0.79 & 0.81 & 0.80 & 0.79 & 0.82 & 0.84 \\\hline
\end{tabular}}
\end{table}

\vspace{-2mm}

\section{Community Detection  with Vertex Features}\label{sec:vertexfeature}

\noindent In Section~\ref{algo}, one could notice that the solutions of the base algorithms are used to derive only the feature vector of each vertex. 
This approach is useful for most real networks, where additional information about 
the vertices is unavailable or difficult to obtain~\cite{Leskovec}. 
However, we hypothesize that if vertex-centric features are available along with the network structure, our algorithm can equally be effective to adopt these features. 
In this case, one needs to replace the ensemble-based features with the real features associated with each vertex (i.e., Step~\ref{algo1:ensemble} in Algorithm 1 is not required in this case). 

To show the generic adaptation of \ENCOD~for overlapping community detection on networks where vertex-features are available, we set the following experimental framework. 
We acquire a {\em coauthorship network}\footnote{\small This network is different from the one mentioned in Table~\ref{dataset}.} 
of Computer Science domains, used in~\cite{ChakrabortySTGM13}, where vertices are authors and edges represent collaborations, with at least one paper written by the adjacent authors. 
There are $24$ research areas (such as Algorithms, Data Mining, etc.), and each paper is annotated with one or more research areas.
The (overlapping) ground-truth communities are marked by the publication venues (conferences/journals). 
The network contains $103,677$ vertices, $352,183$ edges and $1,705$ communities. 

To obtain vertex features, we assign each author $a$ with a feature vector $F_a$ of size $24$ where $F_a(i)$ denotes the fraction of papers published by $a$ in the $i^{th}$ research area.
We run all seven competing algorithms (see Section~\ref{sec:setup}) along with modified non-ensemble based \ENCOD~(we call it \MENCOD) and detect the underlying community structure. 
We further consider SPAEM~\cite{Zhang2015}, a joint probabilistic model to detect communities by combining node attributes and topological structure, as another baseline. 
Table~\ref{result_new} shows the performance of all the competing algorithms along with \ENCOD{} and \MENCOD. We observe that \MENCOD{} is superior to all other competing algorithms (followed by \ENCOD{}).

\begin{table}[tb]
 \centering
 \caption{{\color{black}Performance of the competing algorithms on coauthorship network with vertex feature (O=OSLOM, C=COPRA, S=SLPA, M=MOSES, E=EAGLE, B=BIGCLAM, I=IEDC,  PV = PVOC, ME=MEDOC, SP = SPAEM).}}
 \label{result_new}
 \vspace{-2mm}
  \scalebox{0.8}{
 \begin{tabular}{|l|c|c|c|c|c|c|c|c|c|c|c|c|}
 \hline
	  			& O 	   & C 	  & S    & M    & E    & B  & I  & PV & ME & SP & \ENCOD & \MENCOD \\ \hline
F-Sc 			& 0.78 & 0.76 & 0.81 & 0.78 & 0.75 & 0.79 & 0.73 & 0.74 & 0.76 & 0.80 & {\bf 0.83} & {\bf 0.84} \\  
$\Omega$ 	& 0.76 & 0.80 & 0.83 & 0.78 & 0.78 & 0.76 & 0.76 &0.73  &0.74 & 0.84 & {\bf 0.86} & {\bf 0.87} \\  
ONMI 			& 0.80 & 0.81 & 0.83 & 0.81 & 0.75 & 0.79 & 0.80 &0.76 & 0.79 & 0.81 & {\bf 0.86} & {\bf 0.88} \\ \hline
 \end{tabular}}
 \vspace{-5mm}
\end{table}

\section{Conclusion}

We proposed \ENCOD, the second ensemble algorithm for overlapping community detection by leveraging the outputs of many disjoint CD algorithms. 
We examined the dependencies of our algorithm on different parameters. 
We showed that \ENCOD~ significantly outperforms several standard overlapping CD algorithms, and can be used for detecting the overlapping community structure of networks whose 
vertices are associated with features.

There is still a lack of proper theoretical explanations to justify the superiority of network-based ensemble approaches. 
Also, since all base CD algorithms are not equally important in terms of community assignment for vertices (as observed in Section~\ref{sub:base-algo-effect}), it might be possible to retain only the most important base CD algorithms through a correlation study or a machine learning-based approach. This could also potentially reduce the runtime cost of the algorithm. We plan to explore this possibility in future.


\section*{Acknowledgement}
The authors thank the anonymous reviewers and the editor for their constructive suggestions that helped to improve the paper.
The first author would like to acknowledge the support of Ramanujan Faculty Fellowhsip and the Infosys Center for Artificial Intelligence, IIIT Delhi, India.








\end{document}